\documentclass[10pt]{article}
\usepackage{setspace} 
\usepackage{lineno} 
\usepackage{wrapfig}

\usepackage{graphicx}
\usepackage{amssymb}
\usepackage{amsmath}
\usepackage{epstopdf}

\usepackage{natbib}

\usepackage{sidecap} 
\usepackage{color}
\usepackage{amsmath}
\usepackage{mathabx}
\usepackage{bbold}

\newcommand{\rd}{{\rm d}}

\renewcommand{\vec}[1]{{\ensuremath{ \boldsymbol{ \bf #1}}}}
\newcommand{\vr}{\vec{r}}
\newcommand{\vw}{\vec{w}}

\newcommand{\vC}{\vec{C}}

\newcommand{\vCd}{\vec{C}_{\rm diff}}
\newcommand{\vCp}{\vec{C}_{\rm pref}}
\newcommand{\vCn}{\vec{C}_{\rm null}}
\newcommand{\cp}{c_{\rm pref}}

\newcommand{\cd}{c_{\rm diff}}

\newcommand{\rD}{r^{\rm D}}
\newcommand{\pd}{{\hat{p}}}

\newcommand{\cvec}[2]{ \begin{pmatrix} #1 \\ #2 \end{pmatrix} }
\newcommand{\vone}{\vec{1}}

\title{A note on choice and detect probabilities in the presence of choice bias}

\author{
Ralf M.~Haefner\thanks{ralf.haefner@gmail.com}\\
\small Brain \& Cognitive Sciences, University of Rochester, NY}

\begin{document}

\maketitle

\begin{abstract}
Recently we have presented the analytical relationship between choice probabilities, noise correlations and read-out weights in the classical feedforward decision-making framework (\cite{Haefner2013}). The derivation assumed that behavioral reports are distributed evenly between the two possible choices. This assumption is often violated in empirical data Ð especially when computing so-called Ôgrand CPsÕ combining data across stimulus conditions. Here, we extend our analytical results to situations when subjects show clear biases towards one choice over the other, e.g. in non-zero signal conditions. Importantly, this also extends our results from discrimination tasks to detection tasks and detect probabilities for which much empirical data is available.
We find that CPs and DPs depend monotonously on the fraction, $\pd$, of choices assigned to the more likely option: CPs and DPs are smallest for $\pd=0.5$ and increase as $\pd$ increases, i.e. as the data deviates from the ideal, zero-signal, unbiased scenario. While this deviation is small, our results suggest a) an empirical test for the feedforward framework and b) a way in which to correct choice probability and detect probability measurements before combining different stimulus conditions to increase signal/noise.
\end{abstract}  


\section{Introduction}

Understanding the contribution of sensory neurons to perceptual decisions has been a long-standing goal of systems neuroscience (\cite{Parker1998}). One popular approach to answer this question has been to compute the correlation between individual neuronal variability of those sensory neurons and the choice of the subject in perceptual decision-making tasks. In discrimination tasks, these correlations are quantified as 'choice probability' (CP), and in detection tasks, as 'detect probability' (DP) -- reviewed in \cite{Nienborg2012}. Unfortunately, the interpretation of CPs and DPs is complicated by the fact that they not only depend on how individual neurons are being read-out, but also on how these neurons are correlated with each other (\cite{Shadlen1996}). Recently, we presented the analytical relationship between CPs, read-out weights and noise correlations for the case of a two-alternatives discrimination task assuming that the subject's decisions were unbiased. Here, we extend this work to include biased decision-making when trials are not evenly split between the two choice, and to detection tasks.

\section{Results}

We assume a linear read-out of the responses $\vr\equiv (r_1,r_2,..,r_n)$ of $n$ sensory neurons by a hypothetical decision neuron $\rD=\vw^\top\vr\equiv\sum_{i=1}^n w_i r_i$. We further assume that a decision is made in favor of either choice 1 (of 2 choices, in a discrimination task), or in favor of 'detect' or 'ignore' (in a detection task) when the activity of the decision neuron $\rD$ exceeds a given threshold $\theta$: choice 1 if $\rD>\theta$ and choice 2 if $\rD\le\theta$.
In this mathematical framework discrimination tasks and detection tasks are equivalent and, for simplicity, in the following our notation will use 'choice 1' and 'choice 2' only. Equally, we will use CP to denote both CPs and DPs. We further assume that the response of each neuron is normally distributed around a mean given by its tuning curve, $f_i(s)$ for a particular stimulus value $s$: $r_i=f_i(s)+\eta_i$ where $\vec{\eta}\sim\mathcal{N}(0,\vec{C})$ with covariance matrix $\vec{C}$. 

This allows us to compute the choice-triggered response distribution for each neuron, $p(r_i|\rD>\theta)$ and $p(r_i|\rD<\theta)$, and the choice-triggered average (CTA), i.e. the difference in the mean responses preceding choice 1 and choice 2, respectively (Methods):
\begin{align*}
\Delta_{\rm choice}\vr	&:=	\langle p(r_i|\rD>\theta)\rangle-\langle p(r_i|\rD<\theta)\rangle\\
	&=	\frac{1}{\pd\left(1-\pd\right)}\sqrt{\frac{1}{2\pi}}\frac{\vC\vw}{\sqrt{\vw^\top\vC\vw}}\exp\left\{-\frac{1}{2}\left[\Phi^{-1}\left(\pd\right)\right]^2\right\}\\	
\text{with}\;\pd	&=	\Phi\left[\frac{\vw^\top\vec{f}(s)-\theta}{\sqrt{\vw^\top\vec{C}\vw}}: 0 ; 1\right].
\end{align*}
where $\pd$ is the probability with which the subject chooses choice 1 over choice 2.

CPs can be computed as
\begin{align*}
{\rm CP}_i	&=	\int_{-\infty}^\infty\rd x\, p(r_i=x|\rD<\theta) \int_{x}^\infty\rd y\, p(r_i=y|\rD>\theta)
\end{align*}
For $\pd=0.5$ this integral is identical to that for CPs in the unbiased case. Hypothesizing an equivalent relationship between CTA and CP for the general case (\cite{Haefner2013}), suggests this approximate formula:
\begin{align}
{\rm CP}_i	&\approx	\frac{1}{2}+\frac{\sqrt{2}}{\pi}\frac{(\vC\vw)_i}{\sqrt{C_{ii}\vw^\top\vC\vw}}\frac{\exp\left\{-\frac{1}{2}\left[\Phi^{-1}\left(\pd\right)\right]^2\right\}}{4\pd\left(1-\pd\right)}.
\label{eq_CP}
\end{align}
In simulations we have confirmed that the error induced by the approximation is below 0.5\% for $\pd\le0.9$, which is more than a magnitude smaller than current empirical error bars.

Interestingly, this approximation becomes better as $\pd$ deviates from 0.5 and is therefore better  than the first-order approximation presented for unbiased CPs in \cite{Haefner2013}. This is presumably due to the fact that the choice-triggered distributions, $p(\vr|\rD<\theta)$ and $p(\vr|\rD>\theta)$, are maximally non-Gaussian for the balanced case of $\pd=0.5$ and become progressively more Gaussian as $\pd$ deviates from $0.5$, and that for the Gaussian case the CTA--CP relationship is linear.

\begin{figure}\begin{center}
  \includegraphics[width=7cm]{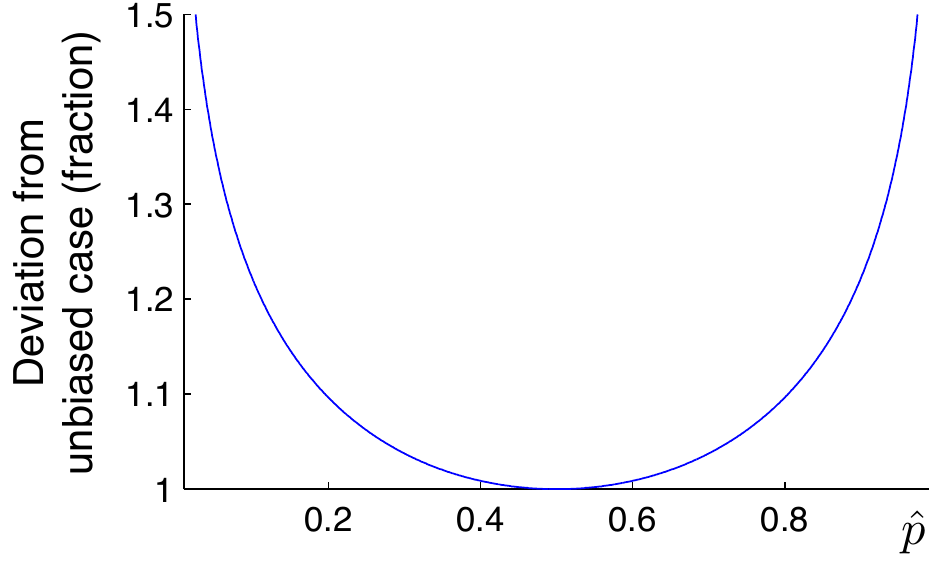}
\caption{Ratio between actual CPs and those for the balanced case of an even number of choices. }
\label{fig_Imbalance}
\end{center}\end{figure}

Figure \ref{fig_Imbalance} plots the bias-dependent term that captures the ratio of the CP in a biased condition to the CP in an unbiased scenario. The bias is quantified by the probability of choice 1, $\pd$, ranging from 0 to 1, with 0.5 being unbiased.

While the read-out strategy for detection tasks is currently unknown, equation (\ref{eq_CP}) implies that for 2-pool models, i.e. read-outs that rely on the difference between the responses of two neuronal pools, the insights gained from discrimination tasks also hold (\cite{Haefner2013}). In particular, in a model in which the averaged responses of two large homogenous neural pools (all neurons have the same response variance) are subtracted from each other, DPs only depend on the difference between the average noise correlation coefficient between neurons in the same pool and the average correlation coefficient between neurons in different pools. This means that DPs in this case depend not on the overall mean of the noise correlations but on their 'structure', i.e. the difference between within and across pool correlations. In contrast, in the equivalent 1-pool model, where the average neuronal activity of a neuronal 'pref' pool is compared to an internal threshold $\theta$, the DP of neurons in the pref pool depends on the average noise correlation within that pool while the DP of other neurons depends on their average correlation with the pref pool neurons. We obtain for large populations (where the diagonal terms can be neglected):
\begin{align*}
\vec{CP}	&	\approx	\frac{1}{2}+\frac{\sqrt{2}}{\pi}\frac{1}{\sqrt{C_{11}\vw^\top\vC\vw}}\begin{pmatrix}\vCp & \vCd \\ \vCd & \vCn\end{pmatrix}\cvec{\vone}{\vec{0}}\frac{\exp\left\{-\frac{1}{2}\left[\Phi^{-1}\left(\pd\right)\right]^2\right\}}{4\pd\left(1-\pd\right)}\quad\text{and therefore}\\
{\rm CP_{pref}}		&\approx	\frac{1}{2}+\frac{\sqrt{2}}{\pi}\sqrt{\cp}
\frac{\exp\left\{-\frac{1}{2}\left[\Phi^{-1}\left(\pd\right)\right]^2\right\}}{4\pd\left(1-\pd\right)}\quad\text{and}\\
{\rm CP_{null}}		&\approx	\frac{1}{2}+\frac{\sqrt{2}}{\pi}\frac{\cd}{\sqrt{\cp}}
\frac{\exp\left\{-\frac{1}{2}\left[\Phi^{-1}\left(\pd\right)\right]^2\right\}}{4\pd\left(1-\pd\right)}
\end{align*}
where $\cp$ is the average correlation between neurons within the pref pool and $\cd$ is the average correlation between neurons that are in different pools.

\section{Discussion}

We have extended the known analytical relationship between choice probabilities, read-out weights and noise correlations in the linear feedforward framework to situations where the subject shows a bias towards one of the two choices. This, more general case, encompasses discrimination tasks where CPs are computed for non-zero-signal trials, and detection tasks and detect probabilities. While this generalization is only approximate, the error due to this approximation is much smaller than current measurement errors. Furthermore, like the approximation previously presented for the balanced case, it allows for an easy intuition into how observable CPs depend on read-out weights, and noise correlations. Notably, it differs from the balanced CP by a factor that only depends on the probability with which the subject chooses choice 1 (or detect) over choice 2 (or miss) -- a value that is easily inferred from the data.

Prior work on the effect of choice bias on CPs (\cite{Kang2012}) has found that CPs are underestimated when grand CPs are computed by z-scoring and combining responses across different signal conditions. This finding is different and not in contradiction with the result presented here, that {\it actual} CPs should increase with choice bias in the feedforward framework.

The dependency of CP on the choice bias that we report has three immediate consequences. One, it suggests an empirical test for the feedforward, linear decoding, decision-making framework that is the basis for the presented relationship. Two, it suggests a way in which to compare CPs measured at different levels of choice bias, for instance in order to combine signal trials and zero-signal trials to increase signal-to-noise in the CP measurements (so-called 'grand CPs', \cite{Nienborg2012}). Third, and most importantly, it extends the benefits of an analytical relationship from discrimination tasks to detection tasks for which much physiological data is available (reviewed in \cite{Nienborg2012}).


\bibliography{../../../../Articles/BibTeX/Paper_CP_DP}
\bibliographystyle{cell} 

\section{Methods}

The choice-triggered response distribution for each neuron is given by:
\begin{align*}
p(r_i|\rD>\theta)	&=	\frac{p(\rD>\theta|r_i)p(r_i)}{p(\rD>\theta)}\\
				&=	\pd^{-1}p(\rD>\theta|r_i)p(r_i)\\
p(r_i|\rD<\theta)	&=	\frac{p(\rD<\theta|r_i)p(r_i)}{p(\rD<\theta)}\\
				&=	(1-\pd)^{-1}p(\rD<\theta|r_i)p(r_i)
\end{align*}
where $\pd=1-\Phi[\theta: \vw^\top\vec{f}(s); \vw^\top\vec{C}\vw]=\Phi[\vw^\top\vec{f}(s)-\theta: 0 ; \vw^\top\vec{C}\vw]$ is the fraction of trials during which the subject reported choice 1. $\pd=0.5$ for the unbiased case. We defined $\phi(r_i:f_i(s);C_{ii})$ to be the Gaussian probability density over $r_i$ with mean $f_i(s)$ and variance $C_{ii}$. Equivalently, $\Phi$ is defined to be the cumulative Gaussian. It follows:
\begin{align*}
p(r_i|\rD>\theta)	&=	\pd^{-1}\phi(r_i:f_i(s);\vec{C}_{ii})\\
			&{}	\left\{1-\Phi\left[\theta : \vw^\top\vec{f}(s)+\frac{(\vC\vw)_i}{C_{ii}}(r_i-f_i(s)) ; \vw^\top\vC\vw-\frac{(\vC\vw)_i^2}{C_{ii}}\right]\right\}\\
			&=	\frac{1}{\pd\sqrt{C_{ii}}}\phi(\frac{\eta_i}{\sqrt{C_{ii}}}:0;1)
				\Phi\left[\frac{\frac{(\vC\vw)_i}{C_{ii}}\eta_i+\vw^\top\vec{f}(s)-\theta}{\sqrt{\vw^\top\vC\vw-\frac{(\vC\vw)_i^2}{C_{ii}}}}:0;1\right]\quad\text{and}\\
p(r_i|\rD<\theta)	&=	\frac{1}{(1-\pd)\sqrt{C_{ii}}}\phi(\frac{\eta_i}{\sqrt{C_{ii}}}:0;1)
				\Phi\left[\frac{\theta-\vw^\top\vec{f}(s)-\frac{(\vC\vw)_i}{C_{ii}}\eta_i}{\sqrt{\vw^\top\vC\vw-\frac{(\vC\vw)_i^2}{C_{ii}}}}: 0;1\right]
\end{align*}
From that the choice-triggered averages can be computed:
\begin{align*}
\langle r_i\rangle|_{\rD>\theta}		&=	\int_{-\infty}^\infty\rd r_i\, p(r_i|\rD>\theta)\\
	&=	\sqrt{C_{ii}}\int_{-\infty}^\infty\rd\left(\frac{\eta_i}{\sqrt{C_{ii}}}\right)\, p\left[f_i(s)+\frac{\eta_i}{\sqrt{C_{ii}}}|\rD>\theta\right]
\end{align*}
We can solve this integral using $\phi'(x)=-x\phi(x)$ and the following auxiliary relationship:
\begin{align*}
\int_{-\infty}^\infty\rd x\, x\phi(x)\Phi(ax-b)	&=	-\phi(x)\Phi(ax-b)|_{-\infty}^\infty + a\int_{-\infty}^\infty\rd x\, \phi(x)\phi(ax-b)\\
		&=	\frac{a}{2\pi}\int_{-\infty}^\infty\rd x\, \exp\left[-\frac{x^2}{2}-\frac{(ax-b)^2}{2}\right]\\
		&=	\frac{a}{2\pi}\int_{-\infty}^\infty\rd x\, \exp\left[-\frac{1+a^2}{2}\left(x-\frac{ab}{1+a^2}\right)^2\right]\exp\left[\frac{a^2b^2}{2(1+a^2)}-\frac{b^2}{2}\right]\\
		&=	\sqrt{\frac{1}{2\pi}}\frac{a}{\sqrt{1+a^2}}\exp\left[-\left(1-\frac{a^2}{1+a^2}\right)\frac{b^2}{2}\right]
\end{align*}
Defining
\begin{align*}
a	&=	\frac{(\vC\vw)_i}{\sqrt{C_{ii}\vw^\top\vC\vw-(\vC\vw)_i^2}}
	=	\sqrt{\left[\frac{(\vC\vw)_i^2}{C_{ii}\vw^\top\vC\vw}\right]^{-1}-1}^{-1}
\Rightarrow
\frac{a}{\sqrt{1+a^2}}	&=	\frac{(\vC\vw)_i}{\sqrt{C_{ii}\vw^\top\vC\vw}}\quad\text{and}\\
b	&=	\frac{\sqrt{C_{ii}}[\theta-\vw^\top\vec{f}(s)]}{\sqrt{C_{ii}\vw^\top\vC\vw-(\vC\vw)_i^2}}
\end{align*}
we find
\begin{align*}
\langle \eta_i\rangle|_{\rD>\theta}		&=	\sqrt{\frac{1}{2\pi}}\frac{1}{\pd}\frac{(\vC\vw)_i}{\sqrt{\vw^\top\vC\vw}}\exp\left[-\frac{1}{2}\left(1-\frac{(\vC\vw)_i^2}{C_{ii}\vw^\top\vC\vw}\right)\frac{C_{ii}[\theta-\vw^\top\vec{f}(s)]^2}{C_{ii}\vw^\top\vC\vw-(\vC\vw)_i^2}\right]\\
&=	\sqrt{\frac{1}{2\pi}}\frac{1}{\pd}\frac{(\vC\vw)_i}{\sqrt{\vw^\top\vC\vw}}\exp\left[-\frac{1}{2}\frac{[\theta-\vw^\top\vec{f}(s)]^2}{\vw^\top\vC\vw}\right]\\
&=	\sqrt{\frac{1}{2\pi}}\frac{1}{\pd}\frac{(\vC\vw)_i}{\sqrt{\vw^\top\vC\vw}}\exp\left\{-\frac{1}{2}\left[\Phi^{-1}\left(\pd\right)\right]^2\right\}
\end{align*}

Equivalently, we find:
\begin{align*}
\langle \eta_i\rangle|_{\rD<\theta}		&=	\int_{-\infty}^\infty\rd\, \eta_i p(r_i|\rD<\theta)\\
a	&=	-\frac{(\vC\vw)_i}{\sqrt{C_{ii}\vw^\top\vC\vw-(\vC\vw)_i^2}}\\
\Rightarrow
\frac{a}{\sqrt{1+a^2}}	&=	-\frac{(\vC\vw)_i}{\sqrt{C_{ii}\vw^\top\vC\vw}}\\
b	&=	-\frac{C_{ii}[\theta-\vw^\top\vec{f}(s)]}{\sqrt{C_{ii}\vw^\top\vC\vw-(\vC\vw)_i^2}}\\
\langle \eta_i\rangle|_{\rD<\theta}		&=	-\sqrt{\frac{1}{2\pi}}\frac{1}{1-\pd}\frac{(\vC\vw)_i}{\sqrt{\vw^\top\vC\vw}}\exp\left[-\frac{1}{2}\left(1-\frac{(\vC\vw)_i^2}{C_{ii}\vw^\top\vC\vw}\right)\frac{C_{ii}[\theta-\vw^\top\vec{f}(s)]^2}{C_{ii}\vw^\top\vC\vw-(\vC\vw)_i^2}\right]\\
&=	-\sqrt{\frac{1}{2\pi}}\frac{1}{1-\pd}\frac{(\vC\vw)_i}{\sqrt{\vw^\top\vC\vw}}\exp\left\{-\frac{1}{2}\left[\Phi^{-1}\left(\pd\right)\right]^2\right\}
\end{align*}
Hence
\begin{align*}
\Delta_{\rm choice}\vr	&:=	\langle p(r_i|\rD>\theta)\rangle-\langle p(r_i|\rD<\theta)\rangle\\
	&=	\frac{1}{\pd\left(1-\pd\right)}\sqrt{\frac{1}{2\pi}}\frac{\vC\vw}{\sqrt{\vw^\top\vC\vw}}\exp\left\{-\frac{1}{2}\left[\Phi^{-1}\left(\pd\right)\right]^2\right\}\\	
\text{with}\;\pd	&=	\Phi\left[\frac{\vw^\top\vec{f}(s)-\theta}{\sqrt{\vw^\top\vec{C}\vw}}: 0 ; 1\right].
\end{align*}

\end{document}